\documentclass[vecphys]{svmult}

\usepackage{makeidx}         
\usepackage{graphicx}        
\usepackage{multicol}        
\usepackage[bottom]{footmisc}

\makeindex             

\newcommand{\LAOG}{1}
\newcommand{\Cavendish}{2}
\newcommand{\MPIfR}{3}
\newcommand{\CAUP}{4}
\newcommand{\INAFOATo}{5}
\newcommand{\INAFOAR}{6}
\newcommand{\IAGL}{7}
\newcommand{\IfAW}{8}
\newcommand{\AIU}{9}
\newcommand{\CEALETI}{10}
\newcommand{\CRAL}{11}
\newcommand{\FCUL}{12}
\newcommand{\INETI}{13}
\newcommand{\INAFOAA}{14}
\newcommand{\JMMC}{15}
\newcommand{\ESOChile}{16}
\newcommand{\ESOHQ}{17}
\newcommand{\IAC}{18}
\newcommand{\OCA}{19}
\newcommand{\GRAAL}{20}

\begin{document}

\title*{Milli-arcsecond astrophysics with VSI, the VLTI spectro-imager in
  the ELT era} 
\titlerunning{VLTI Spectro-Imager (VSI)}
\author{  
F. Malbet \inst{\LAOG} \and 
D. Buscher\inst{\Cavendish} \and
G. Weigelt\inst{\MPIfR} \and
P. Garcia\inst{\CAUP} \and
M. Gai\inst{\INAFOATo} \and
D. Lorenzetti\inst{\INAFOAR} \and
J. Surdej\inst{\IAGL} \and
J. Hron\inst{\IfAW} \and
R. Neuh\"auser\inst{\AIU} \and
P. Kern\inst{\LAOG} \and
L. Jocou\inst{\LAOG} \and
J.-P. Berger\inst{\LAOG} \and
O. Absil\inst{\LAOG} \and
U. Beckmann\inst{\MPIfR} \and
L. Corcione\inst{\INAFOATo} \and
G. Duvert\inst{\LAOG,\JMMC} \and
M. Filho\inst{\CAUP} \and
P. Labeye\inst{\CEALETI} \and
E. Le Coarer\inst{\LAOG} \and
G. Li Causi\inst{\INAFOAR} \and
J. Lima\inst{\FCUL} \and
K. Perraut\inst{\LAOG} \and
E. Tatulli\inst{\LAOG,\INAFOAA,\JMMC} \and
E. Thi\'ebaut\inst{\CRAL} \and
J. Young\inst{\Cavendish} \and
G. Zins\inst{\LAOG} \and
A. Amorim\inst{\FCUL} \and
B. Aringer\inst{\IfAW} \and
T. Beckert\inst{\MPIfR} \and
M. Benisty\inst{\LAOG} \and
X. Bonfils\inst{\FCUL} \and
A. Chelli\inst{\LAOG,\JMMC} \and
O. Chesneau\inst{\OCA} \and
A. Chiavassa\inst{\GRAAL} \and
R. Corradi\inst{\IAC} \and
M. de Becker\inst{\IAGL} \and
A. Delboulb\'e\inst{\LAOG} \and
G. Duch\^ene\inst{\LAOG} \and
T. Forveille\inst{\LAOG} \and
C. Haniff\inst{\Cavendish} \and
E. Herwats\inst{\LAOG,\IAGL} \and
K.-H.Hofmann\inst{\MPIfR} \and
J.-B. Le Bouquin\inst{\ESOChile} \and
S. Ligori\inst{\INAFOATo} \and
D. Loreggia\inst{\INAFOAR} \and
A. Marconi\inst{\INAFOAA} \and
A. Moitinho\inst{\FCUL} \and
B. Nisini\inst{\INAFOAR} \and
P.-O. Petrucci\inst{\LAOG} \and
J. Rebordao\inst{\INETI} \and
R. Speziali\inst{\INAFOAR} \and
L. Testi\inst{\INAFOAA,\ESOHQ} \and
F. Vitali\inst{\INAFOAR}
}%
\authorrunning{F. Malbet et al.} 
\institute{
Universit\'e J.~Fourier, CNRS, Laboratoire d'Astrophysique de Grenoble,
UMR 5571, BP 53, F-38041 Grenoble cedex 9, France
\and 
Cavendish Laboratory of University of Cambridge, UK
\and
Max-Planck Institute for Radioastronomy, Bonn, Germany
\and
Center for Astrophysics of University of Porto, Portugal
\and
INAF/Osservatorio astrofisico di Torino, Italy
\and
INAF/Osservatorio di Astrofisica di Roma, Italy
\and
Institute of Astrophysics and Geophysics,  Li\`ege, Belgium
\and
Institute of Astrophysics of the university of Wien, Austria
\and
Astrophysical Institute and University Observatory, Jena, Germany
\and
Laboratoire d'\'electronique et de technologie de l'information, Grenoble, France
\and
Centre de Recherche en Astrophysique de Lyon, France
\and
SIM/IDL Faculdade de Ci\^encias da Universidade de Lisboa, Portugal
\and
Instituto Nacional de Engenharia, Tecnologia e Inovacco, Lisboa, Portugal
\and
INAF/Osservatorio di Astrofisica di Arcetri, Italy
\and
Jean-Marie Mariotti Center, CNRS, France
\and
European Southern Observatory, Santiago, Chile
\and
European Southern Observatory Headquarters, Garching, Germany
\and
Instituto de Astrof\'isica de Canarias, Spain
\and
Observatoire de la C\^ote d'Azur, Laboratoire Gemini, Nice, France
\and
Groupe de Recherches en Astronomie et Astrophysique du Languedoc, Montpellier, France
}
\maketitle

\begin{abstract}
  Nowadays, compact sources like surfaces of nearby stars, circumstellar
  environments of stars from early stages to the most evolved ones and 
  surroundings of active galactic nuclei can be investigated at
  milli-arcsecond scales only with the VLT in its interferometric
  mode. We propose a spectro-imager, named VSI (VLTI spectro-imager),
  which is capable to probe these sources both over spatial and spectral
  scales in the near-infrared domain. This instrument will provide
  information complementary to what is obtained at the same time with ALMA
  at different wavelengths and the extreme large telescopes.
\end{abstract}

At the beginning of the 21st century, infrared observations performed
at the milli-arcsecond scale are essential for many astrophysical
investigations either to compare the same physical phenomena at
different wavelengths (like sources already observed with the VLBI or
soon to be observed by ALMA) or to get finer details on observations
carried out with the \emph{Hubble Space Telescope} (HST) or 10-m class
telescopes equipped with adaptive optics.  The astrophysical science
cases at milli-arcsecond scales which cover from planetary physics to
extragalactic studies can only be studied using interferometric
aperture synthesis imaging with several optical telescopes.  In this
respect, the \emph{Very Large Telescope} (VLT) observatory of the
\emph{European Southern Observatory} (ESO) is a unique site world-wide
with $4\times8$-m unit telescopes (UTs), $4\times1.8$-m auxiliary
telescopes (ATs) and all the required infrastructure, in particular
delay lines (DLs), to combine up to 6 telescopes. The \emph{VLT
  Interferometer} (VLTI) infrastructure can be directly compared to
the \emph{Plateau de Bure Interferometer} (PdBI) which combines
$6\times15$-m antenna over 500-m in the millimeter-wave domain. The
quality of the foreseen images can be directly compared to the images
provided by the PdBI. However, the angular resolution of the VLTI is a
few hundred times higher due to the observation at shorter
wavelengths.  The large apertures of the VLTI telescopes and the
availability of fringe tracking allow sensitivity and spectral
resolution to be added to the imaging capability of the VLTI.

In April 2005, at the ESO workshop on \emph{``The power of
  optical/infrared interferometry: recent scientific results and
  second generation VLTI instrumentation''}, two independent teams
have proposed two different concepts for an imaging near-infrared
instrument for the VLTI: BOBCAT \cite{Bobcat} and VITRUV
\cite{Vitruv}. In October 2005, the science cases of these instruments
were approved by the ESO \emph{Science and Technical Committee}. In
January 2006, the two projects merged in order to propose the
\emph{VLTI spectro-imager} (VSI) as a response \cite{VSI-PRO-001} to
the ESO call for phase A proposals for second generation VLTI
instruments. The phase A study ended in September 2007 after an ESO
board review.

\section{VSI overview}
\label{sect:overview}

The VLTI Spectro Imager will provide the ESO community with
spectrally-resolved near-infrared images at angular resolutions down
to 1.1 milliarcsecond and spectral resolutions up to $R=12000$.
Targets as faint as $K=13$ will be imaged without requiring a brighter
nearby reference object; fainter targets can be accessed if a
suitable off-axis reference is available. This unique combination of
high-dynamic-range imaging at high angular resolution and high spectral
resolution for a wide range of targets enables a
scientific programme which will serve a broad user community within
ESO and at the same time provide the opportunity for breakthroughs in
many areas at the forefront of astrophysics.

A great advantage of VSI is that it will provide these new
capabilities while using technologies which have extensively been
tested in the past and while requiring little in terms of new
infrastructure on the VLTI. At the same time, VSI will be capable to
make maximum use of the new infrastructure as it becomes available.
VSI provides the VLTI with an instrument capable of combining up to 8
telescopes, enabling rapid imaging through measurement of up to 28
visibilities in hundreds of wavelength channels within a few minutes.
Operations with less than 8 telescopes is the scope of the first
phases of VSI.  Three development phases are foreseen: VSI4
combining 4 telescopes (UTs or ATs), VSI6 combining 6 telescopes
(4UTS+2ATs or 4ATs+2UTS and eventually 6ATs), and perhaps ultimately, in the
long-run, VSI8 combining 8 telescopes (4UTs+4ATs or
eventually 8ATs). \textbf{The current studies were focused on a
  4-telescope version with an upgrade to a 6-telescope one}. The
instrument contains its own fringe tracker and wavefront control in
order to reduce the constraints on the VLTI infrastructure and
maximize the scientific return.

\section{Science cases for VSI}
\label{sect:science}

The high level specifications of the instrument are derived from
science cases based on the capability to reconstruct for the
milli-arcsecond-resolution images of a wide range of
targets. These science cases are detailed below.
\begin{itemize}
\item \textbf{Formation of stars and planets}. The early evolution of
  stars and the initial conditions for planet formation are determined
  by the interplay between accretion and outflow processes. Due the
  small spatial scales where these processes take place,
  very little is known about the actual physical and chemical
  mechanisms at work.  Interferometric imaging at 1 milli-arcsecond
  spatial resolution will directly probe the regions responsible for
  the bulk of excess continuum emission from these objects, therefore
  constraining the currently highly degenerate models for the spectral
  energy distribution (see \ref{fig:yso}).
  \begin{figure}[t]
    \centering
    \includegraphics[width=0.95\hsize]{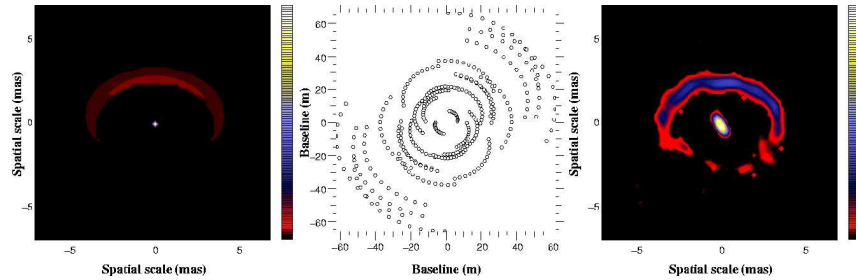}
    \caption{Image reconstruction performed with 6 ATs on a model disk
      around an Herbig Ae star \cite{VSI-PRO-002}. Left: model image;
      middle: coverage of the spatial frequencies; right:
      reconstructed image. The dust structure, the inner dust radius
      and the asymmetry (vertical structure) are well retrieved.
      Relative photometry is reliable (17\% vs 19\% of flux in the
      central star).  }
    \label{fig:yso}
  \end{figure}
  In the emission lines a variety of processes will be probed, in
  particular outflow and accretion magnetospheres.  The inner few AUs
  of planetary systems will also be studied, providing additional
  information on their formation and evolution processes, as well as
  on the physics of extrasolar planets.
\item \textbf{Imaging stellar surfaces}. Optical and near-infared
  imaging instruments provide a powerful means to resolve stellar features
  of the generally patchy surfaces of stars throughout the
  Hertzsprung-Russell diagram.  Optical/infrared interferometry has
  already proved its ability to derive surface structure parameters
  such as limb darkening or other atmospheric parameters.  VSI, as an
  imaging device, is of strong interest to study various specific
  features such as vertical and horizontal temperature profiles and
  abundance inhomogeneities, and to detect their variability as the
  star rotates.  This will provide important keys to address stellar
  activity processes, mass-loss events, magneto-hydrodynamic
  mechanisms, pulsation and stellar evolution.
\item \textbf{Evolved stars, stellar remnants \& stellar winds}. HST
  and ground-based observations revealed that the geometry of young
  and evolved planetary nebulae and related objects (e.g., nebulae
  around symbiotic stars) show an incredible variety of elliptical,
  bi-polar, multi-polar, point-symmetrical, and highly collimated
  (including jets) structures. The proposed mechanisms explaining the
  observed geometries (disks, magnetohydrodynamics collimation and
  binarity) are within the grasp of interferometric imaging at 1~mas
  resolution.  Extreme cases of evolved stars are stellar black holes.
  In microquasars, the stellar black-hole accretes mass from a donor.
  The interest of these systems lies in the small spatial scales and
  high multi-wavelength variability. Milliarcsecond imaging in the
  near-infrared will allow disentangling between dust and jet synchrotron
  emission, comparison of the observed morphology with radio maps and
  correlation of the morphology with the variable X-ray spectral
  states.
\item \textbf{Active Galactic Nuclei \& Supermassive Black Holes}. AGN
  consist of complex systems composed of different interacting parts powered
  by accretion onto the central supermassive black hole. The imaging
  capability will permit the study of the geometry and dust composition of
  the obscuring torus and the testing of radiative transfer models (see
  \ref{fig:agn}).
  \begin{figure}[t]
    \centering
    \includegraphics[width=0.95\hsize]{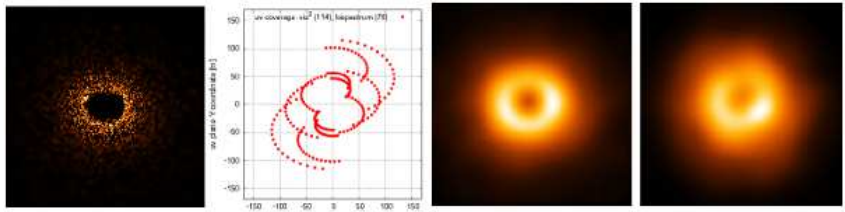}
    \caption{VLTI/VSI image reconstruction simulation performed with 4
      UTs on a model of a clumpy torus at the center of an AGN
      \cite{VLT-SPE-VSI-15870-4337}. Left: model image; middle left:
      coverage of the spatial frequencies; middle right: model image
      convolved with a perfect beam corresponding to the maximum
      spatial resolution; right: image reconstructed from simulated
      VSI data using the Building Block method. }
  \label{fig:agn}
  \end{figure}
  Milli-arcsecond resolution imaging will allow us to probe the
  collimation at the base of the jet and the energy distribution of
  the emitted radiation.  Supermassive black hole masses in nearby
  (active) galaxies can be measured and it will be possible
  to detect general relativistic effects for the stellar orbits closer
  to the galactic center black hole. The wavelength-dependent
  differential-phase variation of broad emission lines will provide
  strong constraints on the size and geometry of the Broad Line Region
  (BLR).  It will then be possible to establish a secure
  size-luminosity relation for the BLR, a fundamental ingredient to
  measure supermassive black hole masses at high redshift.
\end{itemize}

We have shown that this astrophysical program \cite{VSI-PRO-002,
  VLT-SPE-VSI-15870-4337} could provide the premises for a legacy program
at the VLTI. For this goal, the number of telescopes to be
combined should be at least 4, or better 6 to 8 at the VLTI at the
time when the \emph{James Webb Spac*e Telescope} will hammer faint
infrared science ($\sim2013$), when HAWK-I, KMOS, will have hopefully
delivered most of their science, and ALMA will be fully operational.
The competitiveness and uniqueness of the VLT will remain on the high
angular (AO/VLTI) and the high spectral resolution domains. In a
context where the European \emph{Extremely Large Telescope} (ELT) will
start being constructed, then have first light, and, where Paranal
science operations will probably be simplified with less VLT
instruments and an emphasis on survey programs, VSI will take all its
meaning by bringing the VLTI to a legacy mode.

\section{Instrument concept}

The phase A study has led to an instrument concept consisting of:
\begin{itemize}
\item Integrated optics multi-way beam combiners providing
  high-stability visibility and closure-phase measurements on multiple
  baselines;
\item A cooled spectrograph providing resolutions between $R=100$
  and $R=12000$ over the $J$, $H$, or $K$ bands;
\item An integrated high-sensitivity switchable H/K fringe tracker capable of
  real-time cophasing or coherencing of the beams from faint or resolved
  sources;
\item Hardware and software to enable the
  instrument to be aligned, calibrated and operated with minimum
  staff overhead.
\end{itemize}

These features act in synergy to provide a scientific capability which
is a step beyond existing instruments. Compared to the single closure
phase measured by AMBER, the 3 independent closure phases available by
VSI4, the 10 independent closure phases measured by VSI6 and the 21
independent closure phases measured by VSI8 will make true
interferometric imaging, as opposed to simply measuring visibilities,
a routine process at the VLTI.  The capability to cophase on targets
up to $K=10$ will allow long integrations at high spectral resolutions
for large classes of previously inaccessible targets, and the
capability to do self-referenced coherencing on objects as faint as
$K=13$. It will allow imaging of $>$99\% of targets for which no
bright reference is sufficiently close by. VSI will be able to provide
spectrally and spatially-resolved ``image cubes'' for an unprecedented
number of targets at unprecedented resolutions.

\begin{figure*}[t]
  \centering
  \includegraphics[width=\hsize]{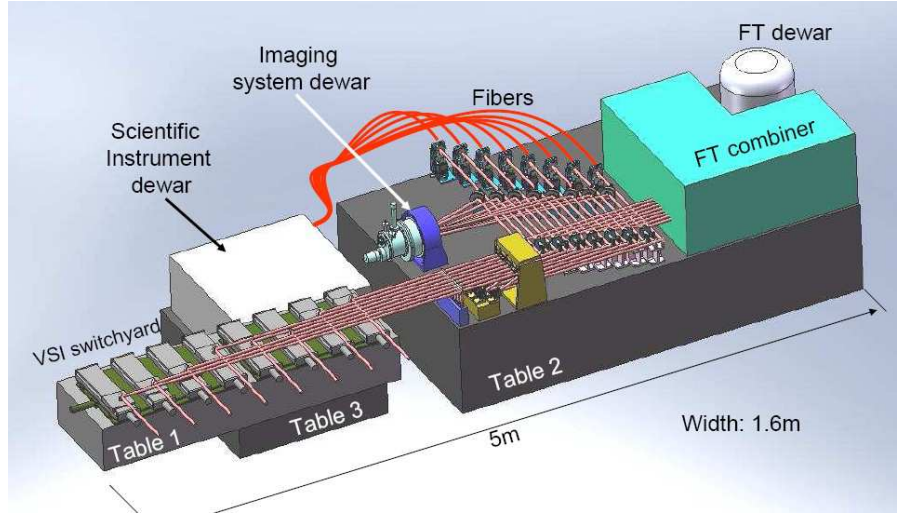}
  \caption{General implementation of the VSI instrument}
  \label{fig:implementation}
\end{figure*}
A system analysis of VSI has allowed the high level specifications of
the system to be defined, the external constraints to be clarified and the
functional analysis to be performed.  The system design
\cite{VLT-SPE-VSI-15870-4339} features 4 main assemblies: the science
instrument (SI), the fringe tracker (FT), the common path (CP) and the
calibration and alignment tools (CAT). The global implementation is
presented in Fig.\ \ref{fig:implementation}.

The optics design of the science instrument features beam combination
using single mode fibers, an integrated optics chip and 4 spectral
resolutions through a cooled spectrograph. The common path includes
low-order adaptive optics (with the current knowledge reduced to only
tip-tilt corrections).  VSI also features an internal fringe tracker.
These servo-loop systems relax the constraints on the VLTI interfaces
by allowing for servo optical path length differences and optimize the
fiber injection of the input beams to the required level. An internal
optical switchyard allows the operator to choose the best
configuration of the VLTI co-phasing scheme in order to perform phase
bootstrapping for the longest baseline on over-resolved objects. Three
infrared science detectors are implemented in the instrument, one
for the Science Instrument, one for the fringe tracker, and one for
the tip-tilt sensor. The instrument features 3 cryogenic vessels.

An important part of the instrument is the control system which
includes several servo-loop controls and management of the observing
software. The science software manages both data processing and image
reconstruction since one of the products of VSI will be a
reconstructed image like for the millimeter-wave interferometers. The
instrument development includes a plan for assembly, integration and
tests in Europe and in Paranal.

An instrument preliminary analysis report
\cite{VLT-TRE-VSI-15870-4341} discusses several important issues such as
the comparison between the integrated optics and bulk optics
solutions, the standard 4- and 6-telescope VLTI array for imaging, the
proposed implementation of M12 mirrors to achieve these configurations
with VSI4 and VSI6, implication of using an heterogeneous array and
analysis of the thermal background.

\section{Requirements on the VLTI infrastructure}

The needs for future VLTI infrastructure can be summarized
\cite{VLT-SPE-VSI-15870-4335} in an increasing order of completeness
as:
\begin{itemize}
\item {Interferometry Supervisor Software (ISS) upgrade}: upgrade from
  4-telescope version to a 6-telescope version allows VSI to use 6
  telescopes of the existing infrastructure for science cases which
  require imaging on a short timescale.
\item {AT5 and AT6:} 2 additional ATs allow the
  VLTI to use VSI in an efficient way without fast reconfiguration of
  the array.
\end{itemize} 
On a longer term, 8T combination at the VLTI could be foreseen but
this is not a VSI priority. In any case, it would require: 
\begin{itemize}
\item {DL7 and DL8:} 2 additional delay lines allow even
  without AT5 and AT6 to use all telescopes on the VLTI (4ATs+4UTs)
  and would be useful for complex imaging of rapidly changing
  sources. 
\item {AT7 and AT8:} could be implemented if DL7 and DL8 are
  procured. Then, the 8T VLTI capability could be exploited only with
  the ATs. 
\end{itemize}

\section{VSI project management}

For VSI4, the management plan \cite{VLT-PLA-VSI-15870-4338} identifies
a total cost of $3986$\,kEuros for hardware and a manpower of 87 FTEs
over 4 years before the commissioning begins. Since the instrument is
designed from the beginning for maximum VLTI capacity, the VSI6
version would cost only $385$\,kEuros and 6 FTEs in addition to the
VSI4 version. It has, however, a stronger impact on the general VLTI
development, especially the ISS.  Based on the letters of intent from
the consortium institutes, we estimated that the consortium can
provide 82 secure full time equivalents (FTEs) and possibly 33
additional FTEs.  On the financial side, many institutes are already
in a negotiation phase with their funding agencies.  A minimum ESO
contribution would be requested for procurements of ESO standard
control boards and possibly for detectors and controllers of the
science and the fringe tracker cameras.

\acknowledgement{The VSI phase A study has benefited from a contract
  by ESO, and support from JRA4 of OPTICON and from CNRS/INSU.}


\printindex
\end{document}